\documentclass[12pt]{article}
\usepackage{epsf}
\usepackage{graphicx}
\usepackage{eepic}
\usepackage{latexsym}
\newcommand{\be}{\begin{equation}}
\newcommand{\ee}{\end{equation}}
\newcommand{\ba}{\begin{array}}
\newcommand{\ea}{\end{array}}
\newcommand{\bqa}{\begin{eqnarray}}
\newcommand{\eqa}{\end{eqnarray}}

\begin{document}

\begin{center}
{\Large\bf An improved study of the kappa resonance and the
non-exotic $s$ wave $\pi K$
  scatterings up to $\sqrt{s}=2.1$GeV of LASS data }
\\[10mm]
{\sc  Z.~Y.~Zhou\footnote{e-mail: zyzhou@pku.edu.cn} and
H.~Q.~Zheng\footnote{e-mail: zhenghq@pku.edu.cn}}
\\[2mm]
{\it  Department of Physics, Peking University, Beijing 100871,
P.~R.~China}
\\[5mm]
\today
\begin{abstract}
We point out that the dispersion relation for the left hand cut
integral presented in one of our previous paper (Nucl. Phys. {\bf
A}733(2004)235) is actually free of subtraction constant, even for
unequal mass elastic scatterings. A new fit to the LASS
data~\cite{Aston} is performed and firm evidence for the existence
of $\kappa$ pole is found.  The correct use of analyticity also put
strong constraints on threshold parameters -- which are found to be
in good agreement with those obtained from chiral theories. We also
determined the pole parameters of  $K_0^*(1430)$ on the second
sheet, and reconfirm the existence of $K_0^*(1950)$ on the third
sheet. We stress that the LASS data do not require them to have the
twin pole structure of a typical Breit--Wigner resonance.
\end{abstract}
\end{center}

\vspace{1cm}
The debate on whether there exists a $\kappa$ meson in the I,J=1/2,0
channel $\pi K$ scattering process has a rather long
history.~\cite{prviouskappa,CP01} In a previous paper~\cite{piK} we
studied the problem on the existence of the $\kappa$ resonance and
have concluded, based on the LASS phase shift data on $\pi K$
scatterings,~\cite{Aston} that the $\kappa$ resonance should exist
if the scattering lengths are close to those predicted by
$\chi$PT.~\cite{Meissner} The conclusion is obtained using a new
parametrization form for the elastic scattering amplitude respecting
unitarity and analyticity.~\cite{piK} It should be emphasized that
one only needs to assume the validity of Mandelstam representation
in order to establish the parametrization form (herewith called the
PKU parametrization form). One of the most remarkable advantages of
the PKU form is that the effects of the  elastic unitarity cut are
dissolved into (2nd sheet) pole contributions and more distant cut
contributions. It can still keep track of contributions from distant
cuts which is of great help in stablizing the fit $\kappa$ pole
location,~\cite{Bugg} comparing with the conventional $K$--matrix
approach. Meanwhile the pole location is not sensitive to the
details of the input left hand cut (in Ref.~\cite{piK} 1--loop
$\chi$PT estimates on the left hand cut was used) and hence the
theoretical uncertainties are severely suppressed. The PKU
parametrization form is found to be sensitive to low lying isolated
singularities not too far away from the elastic threshold and hence
provides a powerful tool especially in exploring those light and
broad resonances.

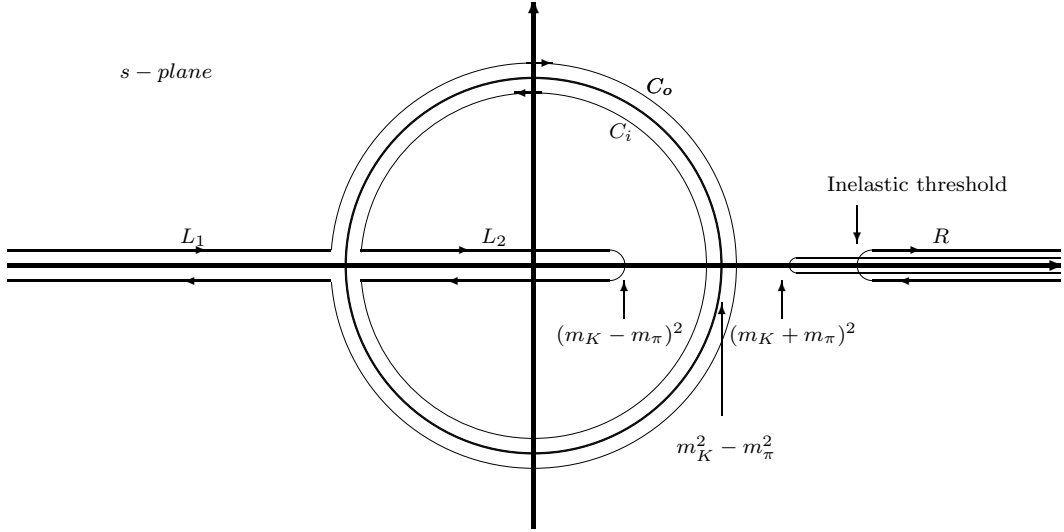
\begin{figure}
\setlength{\unitlength}{1.mm}
\begin{center}%
\begin{picture}(140,70)%
\put(0,35){\vector(1,0){140}}%
\put(0,34.8){\vector(1,0){140}}%
\put(0,35.2){\vector(1,0){140}}%

\put(70,0){\vector(0,1){70}}%
\put(69.8,0){\vector(0,1){70}}%
\put(70.2,0){\vector(0,1){70}}%

\put(70,35){\circle{50}}%
\put(70,35){\circle{49.8}}%
\put(70,35){\circle{50.2}}%

\put(70,35){\arc{54}{3.22}{3.06}}%
\put(70,35){\arc{46}{3.22}{3.06}}%

\put(0,33){\line(1,0){43}}%
\put(0,37){\line(1,0){43}}%
\put(47,33){\line(1,0){33}}%
\put(47,37){\line(1,0){33}}%
\put(115,33){\line(1,0){25}}%
\put(115,37){\line(1,0){25}}%
\put(105,34){\line(1,0){35}}%
\put(105,36){\line(1,0){35}}%

\put(80,35){\oval(4,4)[r]}%
\put(115,35){\oval(4,4)[l]}%
\put(105,35){\oval(2,2)[l]}%

\put(27,33){\vector(-1,0){3.5}}%
\put(23,37){\vector(1,0){3.5}}%
\put(62,33){\vector(-1,0){3.5}}%
\put(58,37){\vector(1,0){3.5}}%
\put(122,33){\vector(-1,0){3.5}}%
\put(118,37){\vector(1,0){3.5}}%

\put(69,62){\vector(1,0){3.5}}%
\put(71,58){\vector(-1,0){3.5}}%

\put(23,38){\scriptsize{$L_1$}} \put(85,58){\scriptsize{$C_o$}}
\put(63,38){\scriptsize{$L_2$}} \put(85,58){\scriptsize{$C_o$}}
\put(123,38){\scriptsize{$R$}} \put(80,52){\scriptsize{$C_i$}}
\put(73,25){\scriptsize{$(m_K-m_{\pi})^2$}}
\put(82,28){\vector(0,1){5}}%
\put(89,10){\scriptsize{$m_K^2-m_{\pi}^2$}}
\put(95,15){\vector(0,1){15}}%
\put(96,25){\scriptsize{$(m_K+m_{\pi})^2$}}
\put(103,28){\vector(0,1){5}}%
\put(109,45){\scriptsize{Inelastic threshold}}
\put(113,43){\vector(0,-1){5}}%
\put(15,60){\scriptsize{$s-plane$}}
\end{picture}%
\end{center}%
\caption{\label{figSS} The left hand cut, circular cut and the
right hand cut of partial wave $\pi K$ scattering amplitudes.}
\end{figure}

The PKU parametrization form for the partial wave elastic
scattering $S$ matrix is,~\cite{piK} \be\label{eq1}
 S^{phy.}=\prod_{i}S^{R_i}\times S^{cut}\ ,
 \ee
 where $S^{R_i}$ denotes the (second sheet) pole contribution, since there are no bound state poles in $\pi K$ scatterings.
 $S^{cut}=e^{2i\rho f(s)}
 $ represents the cut contribution and
$f(s)$ can be expressed as
\bqa\label{fs'}
 f(s) &=& f(s_0)+{(s-s_0)\over 2\pi i}
{\int_C}{{\mathrm{disc}f(z)\over{(z-s)(z-s_0)}}dz} +{(s-s_0)\over
\pi} {\int_{L_1+L_2}}{{\mathrm{Im_L}f(z)\over{(z-s)(z-s_0)}}dz}\nonumber \\%
&+& {(s-s_0)\over \pi}
{\int_R}{{\mathrm{Im_R}f(z)\over{(z-s)(z-s_0)}}dz}\ ,
\eqa
where $L_1$ denotes the left hand cut on the real axis from
-$\infty$ to $-(m_K^2-m_{\pi}^2)$, $L_2$ denotes the left hand cut
from $-(m_K^2-m_{\pi}^2)$ to $(m_K-m_{\pi})^2$, $C$ is the
circular cut, and $R$ denotes cuts from the first inelastic
threshold to
 $+\infty$. See Fig.~\ref{figSS} for explanation.
The discontinuity of $f$ on each cut obeys the following formula,
 \be\label{cutfs}
  \mbox{disc}f(s)=\mbox{disc}\{\frac{1}{2i\rho(s)}\log [S^{phy}(s)]\}\ .
 \ee
  The cut structure of function $f(s)$ as depicted in
Fig.~\ref{figSS} is generated by the partial wave projection of the
un-equal mass elastic scatterings:~\cite{KS}
 \be\label{partialwp} T^I_J(s) = {1\over 32\pi} \int_{-1}^1
d({\mathrm{cos}\theta}) P_J(\mathrm{cos}\theta)T^{I}(s,t,u)\ ,
\ee
 where
  \bqa\label{auxillary}
\mathrm{cos}\theta &=& 1+{{t}\over {2{q_s}^2}}\ ,\nonumber \\%
q_s &=&({(s-(m_K+m_\pi)^2)(s-(m_K-m_\pi)^2)\over{4s}})^{1/2}\ ,\nonumber \\%
u &=& 2m_K^2+2m_\pi^2-s-t\ .
\eqa%
For the convenience of further discussions  Eq.~(\ref{partialwp}) is
rewritten as
 \bqa\label{pwaT} T^I_J(s)
&=& {1\over 32\pi}{1\over 2{q_s}^2} \int_{-4{q_s}^2}^0
dt P_J(1+{{t}\over {2{q_s}^2}})T^{I}(s,t,u)\ .
\eqa%

In Eq.~(\ref{fs'}) there is an arbitrary subtraction constant
$f(s_0)$. In Ref.~\cite{ZHOU05} when discussing $\pi\pi$ scatterings
it was observed that if taking the subtraction at $s=0$ then one can
use the property $T^I_J(0)=const$ to fix the arbitrary subtraction
constant: $f(0)=0$.\footnote{ Fixing the subtraction constant is
very helpful in determining the $\sigma$ pole location in
ref.~\cite{ZHOU05}.} For unequal mass scatterings, however, the
situation becomes more complicated. The simple argument given in
Ref.~\cite{ZHOU05} which led to $f(0)=0$ has to be modified  here.
It is easy to understand this when looking at Eq.~(\ref{pwaT}): when
taking for example the $s\to 0$ limit one has
 \be
  q_s^2\to \frac{(M_K^2-m_\pi^2)^2}{4}s^{-1}\to\infty\ .
   \ee
Therefore in order to reveal the singularity structure of $T^I_J(s)$
when $s\to 0$ one has to evaluate the integral in
Eq.~(\ref{partialwp}) which has to be performed from $t=0$ to
$-\infty$. Hence the asymptotic behavior of $T^I_J(s)$ when $s\to 0$
depends on the asymptotic behavior of the isospin amplitude
$T^I(s,t)$ when $s\to 0$, $t\to \infty$, and the latter is the
energy region governed by Regge model. In Ref.~\cite{Jakob} a
similar problem for $\pi N$ scattering has been studied by assuming
a high energy Regge behavior of $T^I(s,t)$ and it is obtained that
$T^I_J\to s^{-|const|}$ when $s\to 0$, depending on the leading
Regge exchanges in  crossed channels. Nevertheless it is not clear
whether the low partial wave projections of Regge amplitudes should
be totally trustworthy~\cite{collins} and therefore we will not
adopt such an analysis here. We simply point out here that if
assuming the polynomial boundedness of the isospin amplitude:
$|T^I(s,t)|<|t|^n$ when $|t|\to \infty$ and $s$ fixed (which is a
consequence of Mandelstam analyticity), then $T^{I}_J(0)$ will be
singular at most of order $s^{-n}$.\footnote{This statement can be
proved by simply borrowing the method of Ref.~\cite{Jakob}. See
appendix.}
 The latter is enough for the purpose to fix the subtraction constant in the present problem.
 To see this, we evaluate the
asymptotic behavior of \be\label{asyms}
\frac{1}{2i\rho(s)}\log(1+2i\rho(s)T^{phy}(s))
=\frac{1}{2i\rho(s)}\sum_i\log(S^{R_i}(s))+f(s)\ .
 \ee
When $s \rightarrow 0$,   $\rho(s)=2q_s/\sqrt{s}\sim s^{-1}$. Now
if  $T^{phy}(s)\sim O(s^{-n})$, the left hand side vanishes when
$s \rightarrow 0$, and it leads to
\be \label{asy1} f(0)=0\ ,\\
\ee on the $r.h.s.$. The condition Eq.~(\ref{asy1}) should be true
at least for finite number of poles, as a self-consistency
requirement of our approximation scheme.\footnote{ In other words,
taking $f(0)\neq 0$ in the approximation of taking only finite
number of poles will lead to an undesired essential singularity
for $T^{phy}(s)$ at $s=0$.} Thus the dispersion relation
Eq.~(\ref{fs'})  is simplified as,
 \bqa\label{Ff0} f(s) &=& {s\over 2\pi i}
{\int_C}{{\mathrm{disc}f(z)\over{(z-s)z}}dz}+ {s\over \pi}
{\int_{L_1+L_2}}{{\mathrm{Im_L}f(z)\over{(z-s)z}}dz} +{s\over \pi}
{\int_R}{{\mathrm{Im_R}f(z)\over{(z-s)z}}dz}\ .
 \eqa
 Since there is no bound state pole
 in $\pi K$ scatterings Eq.~(\ref{Ff0}) can be recasted  as a more compact form,
 \bqa\label{Ff1} f(s) &=& {s\over 2\pi i}
{\int_{C_{o}}}{{f(z)\over{(z-s)z}}dz}+ {s\over \pi}
{\int_{L_1}}{{\mathrm{Im_{L}}f(z)\over{(z-s)z}}dz} +{s\over \pi}
{\int_R}{{\mathrm{Im_R}f(z)\over{(z-s)z}}dz}
 \eqa
when $s$ lies outside the circular cut. Different from conventional
dispersion relations, the integrand appeared in Eq.~(\ref{Ff1}) is
logarithmic. This is a remarkable property when we use chiral
perturbation theory results to approximate $S^{phy}$ in
Eq.~(\ref{cutfs}), since the bad high energy behavior of chiral
expansions is severely suppressed. As a consequence, in the I=1/2
channel it is found that~\cite{piK} the physical outputs are not
sensitive to the details of the left hand cut contribution at all.
Similar situation happens in the studies of the I=0, J=0 channel
$\pi\pi$ scattering.~\cite{ZHOU05} What is truly important to
introduce the estimation on the left hand cut when studying the low
lying broad resonance is, as revealed in Ref.~\cite{XZ00}, that the
left hand cut contribution to the phase shift is negative and
concave, hence the $\kappa$ ($\sigma$) pole has to be introduced in
order to saturate the experimental data. The observation that the
background contributions are negative and concave as obtained by
using chiral perturbation theory may be doubtful as the latter does
not work at high energies. Nevertheless we believe that the observed
qualitative behavior of the background contribution has little to do
with the bad high energy behavior of chiral amplitudes. There are
two reasons: first of all, as can be verified from Eqs.~(\ref{Ff1})
and (\ref{cutfs}), as long as $|S^{phy}|>1$, the left hand cut
contribution to the phase shift is always negative and concave. But
elastic scattering amplitudes are dominated by pomeron exchanges at
high energies which naturally satisfy $|S^{phy}|>1$. The second
reason is that,  at moderately high energies on the left side, one
may expect that results from chiral perturbation theory are not as
bad as it behaves on the right hand side. Because there is no
unitarity constraints on the left.  Also the left hand side is
further away from resonance (which is one of the main reason for
chiral perturbation theory to break down) region and on the left
side pole singularities are converted to much mild cut
singularities.\footnote{One may find some useful discussions on this
point in Ref.~\cite{HXZ01}.}

 In Ref.~\cite{piK} a fit to the LASS data up to 1.43 GeV was
performed with seven parameters. Six of them are used by MINUIT: two
for the $\kappa$ pole, two for $K_0^*(1430)$ pole and two scattering
lengths (or equivalently the two subtraction constants), $a^{1/2}_0$
and $a^{3/2}_0$. In addition there is another manually tuned cutoff
parameter
 $\Lambda_L^2$ being responsible for the truncation of
the left hand cut integral, chosen at $\Lambda^2_L\simeq
1.5$GeV$^2$. The cutoff parameters in the I=1/2 channel and in the
I=3/2 channel were chosen to be equal, and the value was fixed for
getting a minimal $\chi^2$ in Eq.~(48) of Ref.~\cite{piK}.
 The two
scattering lengths are free parameters in Ref.~\cite{piK}. In here
according to previous discussions, the scattering lengths are no
longer free. In the following  we present the numerical fit results
with six parameters: two for the $\kappa$ pole, two for the
$K_0^*(1430)$ pole, one cutoff parameter in the  I=1/2 channel and
one in the I=3/2 channel. Other conditions are the same as those led
to Eq.~(48) of Ref.~\cite{piK} for   convenience of comparison, but
remembering that now we use instead Eq.~(\ref{Ff1}). The results are
listed below:
 \bqa\label{resLass}
&&\chi^2_{d.o.f.}=80.3/(60-6)\ ;\nonumber\\
&&M_\kappa=694\pm 53MeV\ ,\,\,\, \Gamma_\kappa=606\pm 59MeV\ ;\nonumber\\
&&M_{K^{II}_0(1430)}=1443\pm 15MeV\ ,\,\,\,
\Gamma_{K^{II}_0(1430)}=199\pm 35MeV\ ;
 \eqa
with $\Lambda_{1/2}^2=-13.6\pm 40.0$GeV$^2$,
$\Lambda_{3/2}^2=-11.4\pm 1.7$Gev$^2$. Also a calculation gives
  \bqa\label{resLassthp}
&&a^{1/2}_0=0.219\pm0.034\ ,\,\,\,a^{3/2}_0 =-0.042\pm 0.002\ .
 \eqa
From the above results we  observe that, first of all,  unlike
Ref.~\cite{piK} where one finds, if taking the scattering lengths to
be totally free, a much better $\chi^2$ fit to the data can be
gained with considerably larger (in magnitude) scattering length
parameters, at the cost of introducing an essential singularity at
$s=0$. The better use of the analyticity property seems to further
suggest that the low energy LASS data, to some extent, conflict
other results based on chiral symmetry and $S$ matrix theory.
Nevertheless the existence of $\kappa$ resonance is undoubted within
the present scheme. For example, if freezing one pole the fit gives
$\chi^2_{d.o.f.}=1055.0/(60-4)$. It is worth emphasizing again that
the present scheme indicates, even using the LASS data, that the
threshold parameters are in good agreement with results obtained
using chiral symmetry and $S$ matrix theory properties, for example,
the
 Roy--Steiner equations~\cite{Buttiker} and also
 the dispersive approach of Ref.~\cite{kruvochenko}.
 The effective range parameters are found to be
$b^{1/2}_0=0.075\pm 0.023$ and $b^{3/2}_0=-0.027$, which also agree
fairly well with the results given in Ref.~\cite{Buttiker}, despite
that the present $b^{3/2}_0$ contains a too small error bar. It is
also worth pointing out that the large error bar of
$\Lambda_{1/2}^2$ indicates that our results are not sensitive to
the details of the left hand cut contribution in the I=1/2 channel.
The qualitative behavior of the background contribution in the
I=1/2, J=0 channel does play a crucial role as discussed previously.
Imagine that if the background contributions in this channel were
estimated to be positive and convex. Then the existence of the
$\kappa$ pole would be severely doubted.

 The exotic I=3/2, J=0
channel deserves more discussions. The numerical fit indicates a
large $\Lambda_{3/2}^2$ with a small error bar and hence one may
doubt that our numerical outputs are not trustworthy since the left
hand cut effects in the large $|s|$ region are simply not calculable
by $\chi$PT. Indeed the underestimated error bars for $a_0^{3/2}$
and $b_0^{3/2}$ may already imply that the chiral estimation on left
cuts encounter problems. We will find the reason why MINUIT chooses
such a large cutoff in the exotic channel a short while later. The
concrete number of $\Lambda_{3/2}^2$ itself is governed by the bad
high energy behavior of chiral amplitudes and should not be
trustworthy, but one may think of it in another way: it is data to
choose an appropriate cutoff value to parameterize itself. In this
way the obtained threshold parameters are found to be in fair
agreement with results obtained from other methods. Notice that here
we manage to fit many data using only a single parameter, and still
manage to get a fairly good result. We also point out that in
Ref.~\cite{piK} very careful analyses are made on the influence of
the exotic channel to the determination of the $\kappa$ pole and the
conclusion is that uncertainties in the exotic channel has only
minor effects on outputs in the non-exotic channel.

\begin{figure}%
\begin{center}%
\mbox{\epsfxsize=65mm\epsffile{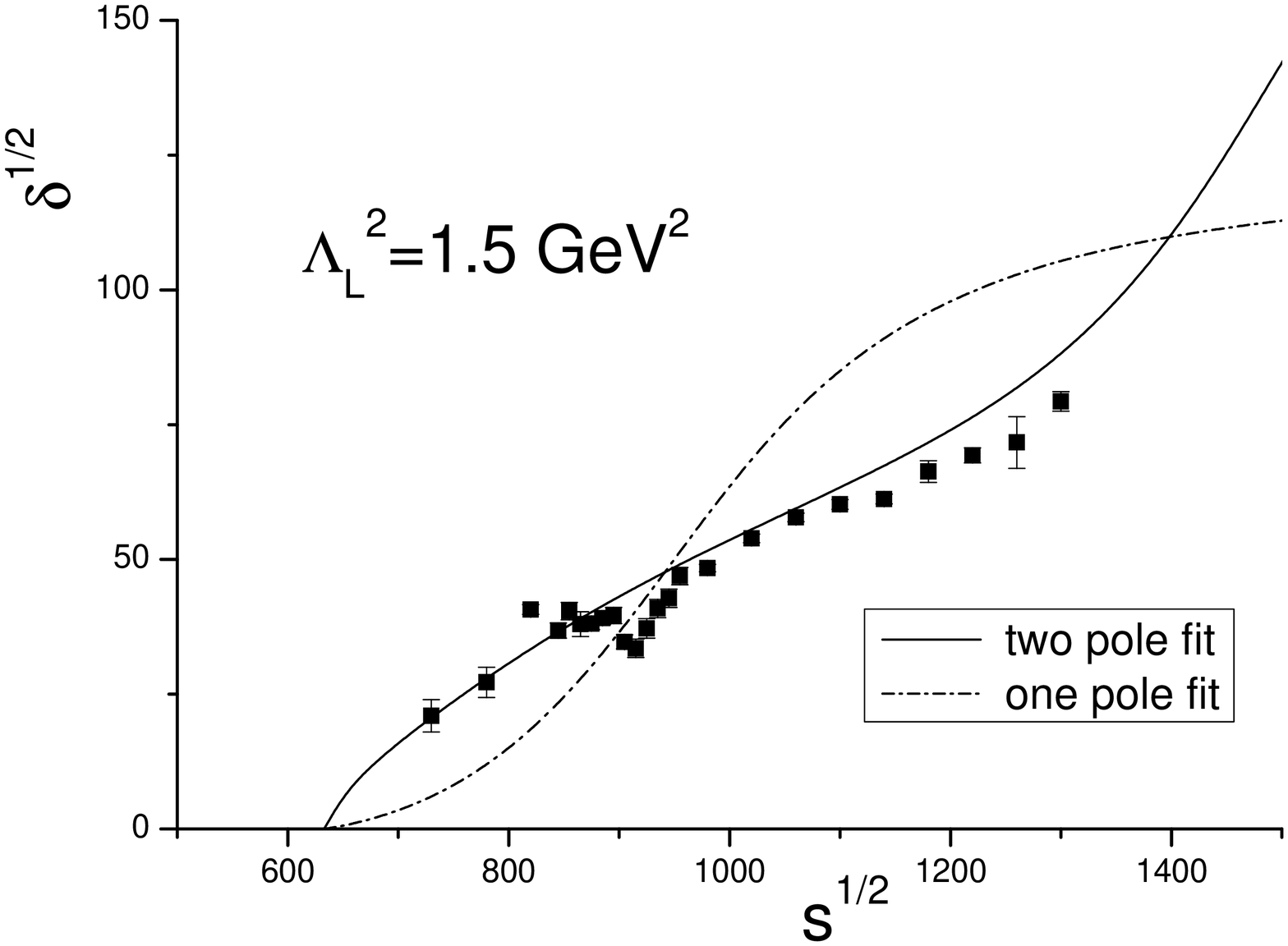}}%
\hspace{.5cm}
\mbox{\epsfxsize=65mm\epsffile{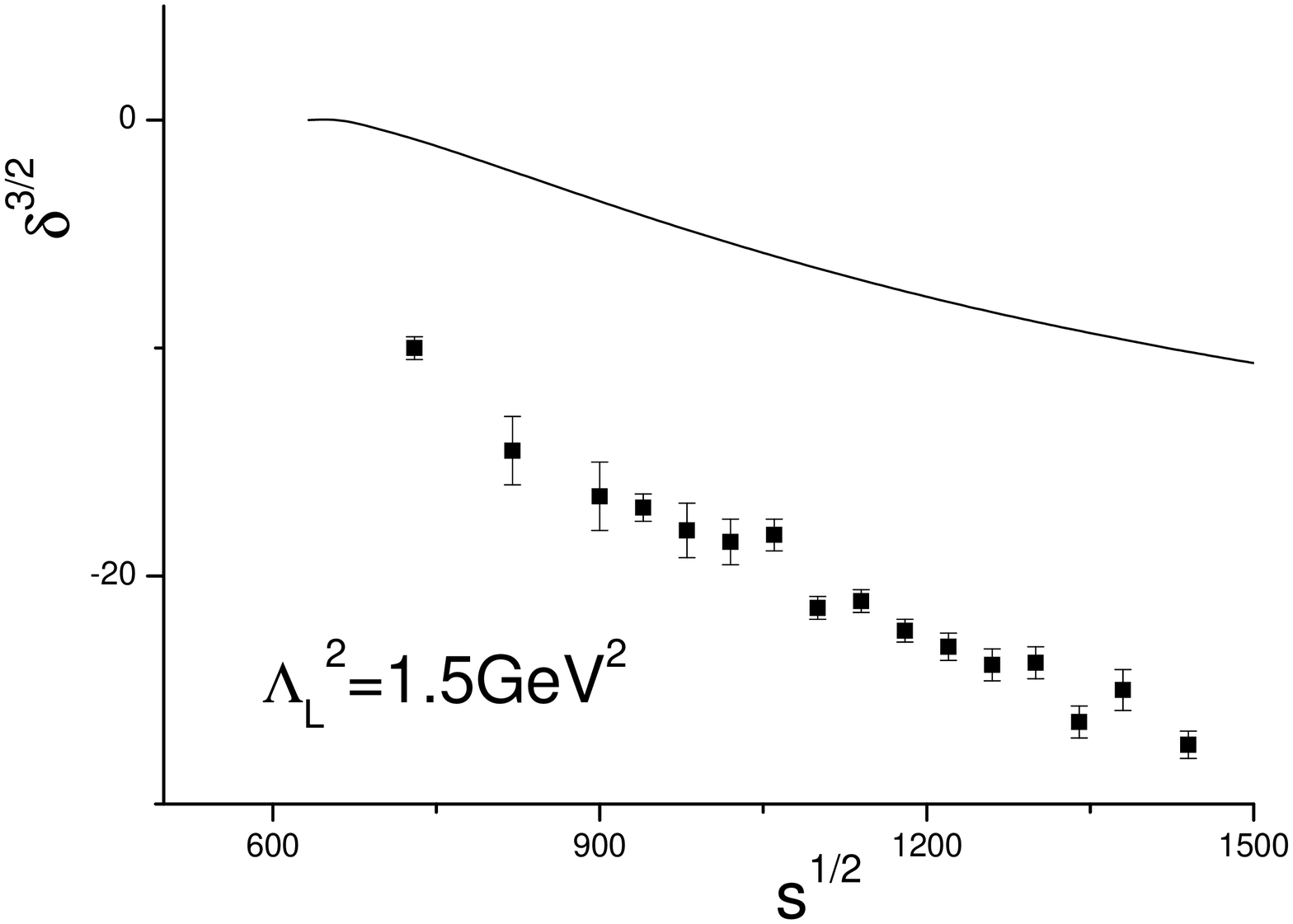}}%
\caption{\label{compareEstabrooks}Phase shifts obtained from the fit
to LASS data up to 1510MeV, to be compared with the data from
Ref.~\cite{Estabrooks}. For $\delta^{1/2}_0$, solid line: two--pole fit, dashed line: one pole fit.}%
\end{center}
\end{figure}

Nevertheless, the appearance of a large $\Lambda_{3/2}^2$  in
Eq.~(\ref{resLass}) may be unpleasant. It is therefore worthwhile to
reexamine in the present scheme how  the $\kappa$ pole relies on the
cutoff parameter in the exotic channel. To do this, we make a test
by fixing the two cutoff parameters both at -1.5GeV$^2$ and fit the
LASS data up to 1510MeV with two poles. The results follow:
 \bqa\label{resLass'}
&&\chi^2_{d.o.f.}=623.6/(68-4)\ ;\nonumber\\
&&M_\kappa=651\pm 20MeV\ ,\,\,\, \Gamma_\kappa=685\pm 13MeV\ ;\nonumber\\
&&M_{K^{II}_0(1430)}=1491\pm 7MeV\ ,\,\,\,
\Gamma_{K^{II}_0(1430)}=346\pm 14MeV\ .
 \eqa
The central value of the scattering lengths are found to be
 $a^{1/2}_0=0.259$, $a^{3/2}_0 =0.006$. Notice that  $a^{3/2}_0$ has a wrong sign here.
 It is not a surprise that the
$\chi^2_{d.o.f.}$ is much worse comparing with Eq.~(\ref{resLass}).
Nevertheless the large $\chi^2_{d.o.f.}$ has little to do with the
I=1/2 channel, rather it is not difficult to ascribe this fault to
the disaster, as it should, happened in the I=3/2 channel. This
excuse is not difficult to accept when comparing the phase shift
curves with the data from Ref.~\cite{Estabrooks}, see
Fig.~\ref{compareEstabrooks}. As has already been emphasized in
Ref.~\cite{piK}, our scheme can rather clearly separate different
channels' effects from the mixed LASS data.
The $K^*_0(1430)$ pole outputs in Eq.~(\ref{resLass'}) should not be
taken seriously since the couple channel effects already become
important at 1500MeV. We will improve our program by incorporating
the inelasticity effects later in this paper. What is really
surprising and remarkable here is that outputs in the I=1/2, J=0
channel are affected very little considering the magnificent change
of $\chi^2_{d.o.f.}$. Furthermore the existence of $\kappa$ can
still be verified by eliminating $\kappa$ in the fit, which results
in $\chi^2_{d.o.f.}=3726/(68-2)$ and $a^{1/2}_0=0.008$, which are
certainly un-acceptable.

It is also noticed that $a^{3/2}_0$ is small and $positive$ when
taking $\Lambda_L^2=1.5$GeV$^2$. A careful check reveals that in the
I=3/2,J=0 channel the circular cut contributes   $positively$ to the
scattering length, contrary to what happens in the I=1/2, J=0
channel. It is now clear why in Eq.~(\ref{resLass})
$\Lambda_{3/2}^2$ takes a large value: the left hand cut on the real
axis has to take a large cutoff to counteract the positive
contribution from the circular cut. Whether a NNLO $\chi$PT
calculation to the background contribution can reduce such a large
$\Lambda_{3/2}^2$ remains open. But from the experience in $\pi\pi$
scatterings one may bet that certain improvement is within
expectation.\footnote{In Fig.~1 of Ref.~\cite{ZHOU05} one finds that
the NNLO correction to the background contribution is towards the
right direction in reducing the magnitude of the cutoff parameter in
the exotic channel. } Furthermore the present investigation strongly
suggests that the high energy contributions are non-negligible in
the exotic channel, irrespective to the fact that it cannot be
reliably estimated by $\chi$PT. Before any further progress can be
made on the exotic channel it is desirable in the following
discussions to proceed along with the philosophy adopted when
deriving Eq.~(\ref{resLass}). That is to think the large
$\Lambda_{3/2}^2$ provides an effective parametrization and is
appropriate to separate the exotic channel contribution from  LASS
data, since the threshold parameters extracted are in agreement with
other theoretical estimates, and especially, the $\kappa$ pole is
influenced rather little with respect to the uncertainties in the
exotic channel.

The Eq.~(\ref{resLass}) indicates a $\kappa$ pole mass at about
$M_\kappa\simeq 700$MeV, $\Gamma_\kappa\simeq 600$MeV, which agrees
with Ref.~\cite{JOP}, and is closer to the recent results from
production experiments,~\cite{kappa-exp} comparing with Eq.~(48) of
Ref.~\cite{piK}. Also it is not strange that the present results are
fully compatible with the results of Eqs.~(51), (52) and (54) of
Ref.~\cite{piK} since the latter are obtained by imposing the
constraints of scattering lengths from chiral estimates.

In the following we also try to extend the data fit range from
1.43GeV to 2.1GeV. The parametrization  of inelasticity plays an
important role in the attempt to include the region above the
inelastic threshold:
 \bqa\label{IMR}
&&\mathrm{Im}_{R}f(s)=-{1\over
{2\rho(s)}}\log|S^{phy}(s)|\nonumber\\
&&=
-\frac{1}{4\rho}\log\left[1-4\rho\mathrm{Im}_{R}T+4\rho^2|T(s)|^2\right]\,\nonumber\\
&&=
-\frac{1}{4\rho}\log\left[1-4\rho(\sum_n\rho_n|T_{1n}(s)|^2+\cdots
)\right]\ .
\eqa%
The approximation we adopt for the right hand cut are from
Ref.~\cite{WJJ}. When third-sheet or higher sheet pole
contributions dominate inelasticity, we have\cite{collins}
  \bqa\label{inelasticT}T_{1n}=\frac1{\sqrt{\rho_1(s)\rho_n(s)}}
  \sum_{r}{ {\cal M}_r(\Gamma_{r1}\Gamma_{rn})^
  {\frac{1}{2}}\over {\cal M}_r^2-s-i{\cal M}_r \Gamma_r}\, ,
  \eqa with $\Gamma_{rn}$ the partial width, $\Gamma_{r}$ the total
  width.

The $K \pi$ scatterings in both $I=1/2$ and $I=3/2$ channels are
inelastic when $K \eta'$ channel opens above roughly 1.4
GeV.\footnote{ The inelasticity in $K \eta$ channel is  quite small
according to $SU(3)$ symmetry~\cite{CP01}, as also verified by
experiments.} Actually, a $K_0^*(1950)$ resonance was observed by
LASS experiment~\cite{Aston}. Assuming inelasticity in I=1/2 channel
is contributed solely from $K_0^*(1950)$, the right hand cut
integral is evaluated by using Eq.~(\ref{IMR}) and
Eq.~(\ref{inelasticT}). As for the I=3/2 channel, it is thought to
be elastic below 1.5 GeV according to its small cross section. Above
1.5GeV the phase shift begin to rise slowly according to the
analysis given in Ref.~\cite{Estabrooks}. It should be noticed that
within the present approximation scheme it is impossible to explain
such a rise of phase shift. Because the J=0, I=3/2 channel is exotic
and what we have to use  in Eq.~(\ref{inelasticT}) is the resonance
dominance approximation.  What we assume here is that the
inelasticity in the exotic channel solely comes from the left hand
cut approximated by $\chi$PT results and the contribution coming
from the right hand cut is negligible. This may be considered as a
drawback of the present approach but hopefully  its influence to the
determination of the $K_0(1950)$ pole location is not large.

There are two solutions of data in Ref.~\cite{Aston}. They are the
same below 1.8 GeV, but different above 1.8 GeV due to Barrelet
ambiguity. The data in the energy region roughly between 1.6 to
1.8 GeV should not be trusted. The reason is very
simple,\cite{JOP}
 \bqa\label{a0phi0}
  a_0e^{i\phi_0}&=&{1\over 4}[(2\eta_{1/2}\sin2\delta_{1/2}+\eta_{3/2}
  \sin2\delta_{3/2})\ ;\nonumber\\
  &+& i (3-2\eta_{1/2}\cos2\delta_{1/2}-\eta_{3/2}\cos2\delta_{3/2})]
\eqa where $a_0$ and $\phi_0$ are the quantities  LASS experimental
group measured. Unitarity demands that the imaginary part of the
$r.h.s.$ of the above equation be positive definite, so it is
impossible for $\phi_0$  to be larger than $180^\circ$.  For the
same reason, we give up fit to the solution B of Ref.~\cite{Aston}.

In the J=0, I=1/2 channel there are contributions from the left hand
cut;  3 second sheet poles, $\kappa$, $K^{II}_0(1430)$ and one to be
determined; two third sheet poles $K_0^{III}(1430)$ and
$K^{III}_0(1950)$ through the right hand cut integrals. The fit
results up to 2.1GeV are shown below,
\begin{figure}%
\begin{center}%
\mbox{\epsfxsize=110mm\epsffile{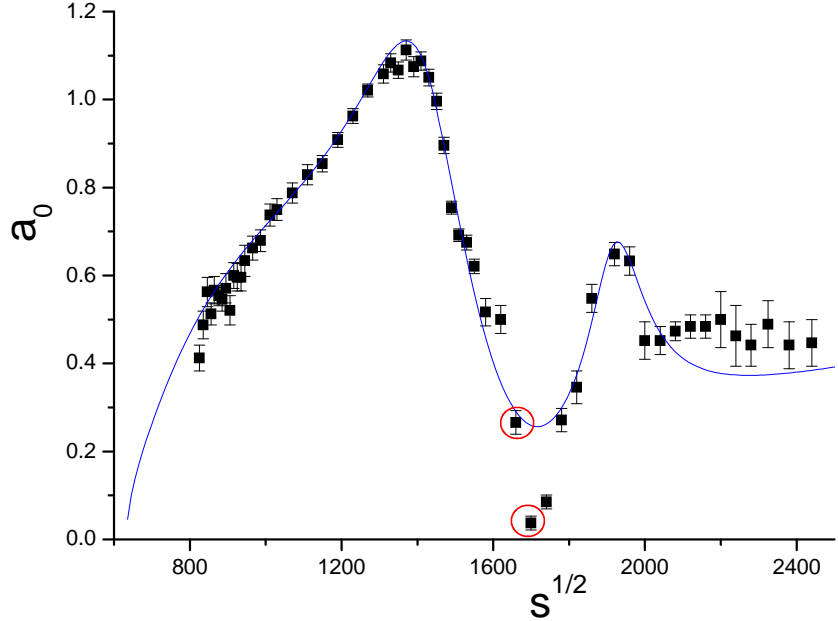}}%
\caption{\label{isospin-3} Fit to $a_0$ of LASS data up to 2.1GeV
(data points encircled  violate unitarity and are not used in the fit).}%
\end{center}%
\end{figure}%
\begin{figure}%
\begin{center}%
\mbox{\epsfxsize=110mm\epsffile{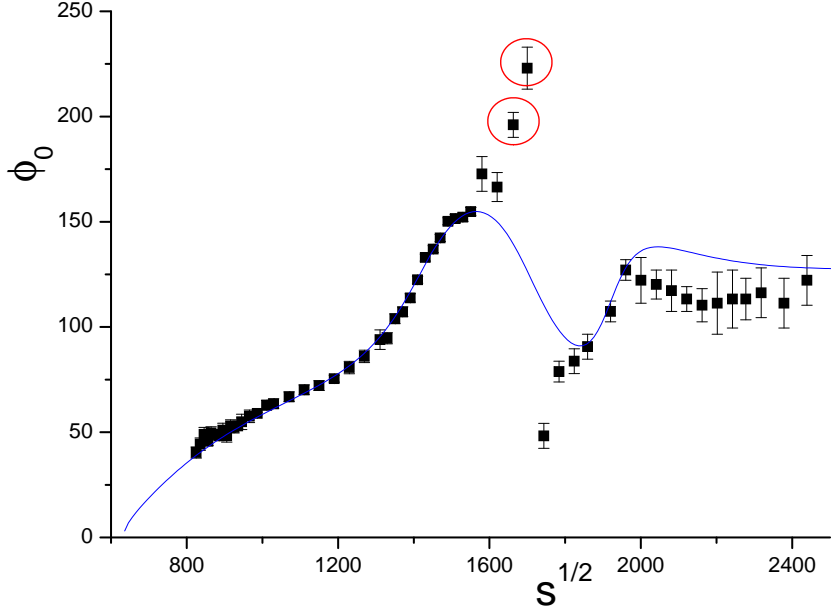}}%
\caption{\label{isospin-3} Fit to $\phi_0$ of LASS data up to
2.1GeV
(data points marked by red circles violate unitarity)}%
\end{center}%
\end{figure}%
\bqa\label{inelasticLassto2100}
&&\chi^2_{d.o.f.}= 229.5/(96-14)\ ;\nonumber\\
&&M_\kappa=649\pm 63MeV\ ,\,\,\, \Gamma_\kappa=602\pm 22MeV\ ;\nonumber\\
&&M_{K^{II}_0(1430)}=1435\pm 6MeV\ ,\,\,\, \Gamma_{K^{II}_0(1430)}=288\pm 22MeV\;\nonumber\\
&&M_{K_0^*(1950)}=1917\pm 12MeV\ ,\,\,\, \Gamma_{K_0^*(1950)}^{tot}=145\pm 38MeV\ ;\nonumber\\
&&\Gamma_{K_0^*(1950)}^{K \pi}=87\pm 14MeV\ .
 \eqa
 Notice that the encircled data points
as depicted in Figs.~2 and 3 are subtracted when making the fit. The
threshold parameters are also consistent with the previous results,
though $a_0^{1/2}$ and $b_0^{1/2}$ are increased by roughly
1$\sigma$ error bar. Not listed in above results are
$K_0^{III}(1430)$
 and the additional second sheet pole (which simulates the possible second sheet $K_0^*(1950)$ pole)
 except the $\kappa$ and
 $K_0^{II}(1430)$.
  Because it is found that these two
 poles (contributing 5 additional  parameters together) only reduce the total $\chi^2$
 roughly by 2, hence their existence is not apparent at
 all.
 Without them in the fit only makes tiny changes to other physical outputs.
 We will discuss again related topics later in this paper.
  Before that we want to check the cutoff dependence of the above
 outputs. If fixing the two cutoff parameters at for example
 $\Lambda_L^2=1.5$GeV$^2$ and let other conditions being the same as
 in deriving Eq.~(\ref{inelasticLassto2100}), then one finds
$\chi^2_{d.o.f.}=777/(96-12)$, and the $\kappa$ pole's  mass and
width are both further decreased by roughly 100MeV comparing with
Eq.~(\ref{inelasticLassto2100}) but with ridiculously small error
bars. Hence the results are un-trustworthy. For the case without
$\kappa$ the output is rather similar at low energies to the
situation of the one pole fit as depicted by
Fig.~\ref{compareEstabrooks}, which is apparently un-physical.

 The $\kappa$ resonance pole position determined in Eq.~(\ref{inelasticLassto2100})
 agrees
 within 1$\sigma$ with that of Eq.~(\ref{resLass}).
  The third sheet $K_0^*(1950)$'s mass, total width and its
partial decay width into $K\pi$ are found in good agreement with the
results obtained by the LASS Collaboration~\cite{Aston} and that of
Anisovich and Sarentsev~\cite{anisovich}. The major difference
between Eq.~(\ref{inelasticLassto2100}) and Eq.~(\ref{resLass}) is
the decay width of $K^{{II}}_0(1430)$. In Eq.~(\ref{resLass}) the
$K^{{II}}_0(1430)$ pole is determined by fitting data only up to
1430MeV and hence the outputs on $K^{{II}}_0(1430)$ parameters given
by Eq.~(\ref{inelasticLassto2100}) may be more preferable.
Furthermore, unlike the situation of $f_2(1270)$~\cite{ZHOU05} where
the twin pole structure of $f_2(1270)$ is clearly identified, here
we are lacking of the separate data of the inelasticity parameter in
the I=1/2 channel. Hence the observation on the absence of
$K^{III}_0(1430)$ from the analysis in
Eq.~(\ref{inelasticLassto2100}) only provides a possibility, but not
a definite conclusion. On the other side, the analysis on the
unitarized scattering amplitudes from resonance chiral lagrangian
model do generate a typical couple channel Breit--Wigner resonance
structure for $K_0^*(1430)$.~\cite{JOP} The PKU parametrization
form, which makes clear distinction between the second sheet pole
and the third sheet pole, is more sensitive to the second sheet pole
rather than the third sheet pole in the absence of the data of
inelasticity parameter. Different from the $K_0^*(1430)$ pole, for
the $K_0(1950)$ resonance on the third sheet, our analysis
definitely confirms its existence. However, in disagreement with
Ref.~\cite{JOP}, we did $not$ find evidences in support of its
counterpart on the second sheet. A narrow second sheet pole around
1950MeV, according to Eq.~(\ref{eq1}), will provide a rapid phase
motion for $\delta$ and also for $\phi_0$ around 1950MeV.
Nevertheless it is seen from fig.~\ref{isospin-3} that the major
part of the enhancement of $\phi_0$ around 1950MeV may already be
explained by $K^{III}_0(1950)$ combined with other background
contributions. If our observation is correct it would suggest that
the $K_0^*(1950)$ may not be a typical Breit--Wigner
resonance.\footnote{Another possibility is that $K^*_0(1950)$
 is still a couple channel Breit--Wigner resonance, but the two poles locate on
 third and fourth sheet, respectively. In the present scheme we are unable to investigate
 such a possibility, since the fourth sheet pole does not enter the approximate
 parametrization form Eq.~(\ref{inelasticT}).} But of course,
more careful analysis on more accurate data is needed to clarify
this issue.

The stability of our fit results on the 3rd sheet $K_0^*(1950)$ pole
can be further investigated. In getting
Eq.~(\ref{inelasticLassto2100}) the fit is performed up to 2.1GeV.
The effects from possible higher resonances were not considered. In
order to check the stability of the results we also made the fit up
to 2.5GeV using the same parameter set as
Eq.~(\ref{inelasticLassto2100}) is obtained, and bare in mind that
the two additional 2nd sheet and 3rd sheet poles now may effectively
simulate effects from higher energies. Indeed, the fit indicates
these additional poles now locate at around 3GeV with sizable error
bars. Of course these additional outputs themselves are of no value
to mention. The major concern is the stability of the $K_0^*(1950)$
pole. It turns out that the rest of the fit results are very similar
to that of Eq.~(\ref{inelasticLassto2100}) except that the total
width of the 3rd sheet $K_0^*(1950)$ pole is now reduced by less
than 20MeV (with the branching ratio almost unchanged). Such a
reduction in total width is acceptable since it's within the error
as given in Eq.~(\ref{inelasticLassto2100}).

To conclude, comparing with Ref.~\cite{piK} the correct use of
analyticity led us to have stronger confidence on the existence of
the $\kappa$ resonance, based on the LASS data.  Scattering length
parameters are also found to be, naturally, in agreement with
chiral theory results. The pole mass and width for the low lying
$\kappa$ resonance are found to be $M_\kappa=694\pm 53MeV$,
$\Gamma_\kappa=606\pm 59MeV$, respectively. The $K_0^*(1430)$ and
$K_0^*(1950)$ resonances are also studied by fitting the LASS data
up to 2.1GeV and the result is given in
Eq.~(\ref{inelasticLassto2100}). Finally, according to the LASS
data we did not find it necessary to introduce the $K_0^*(1430)$
pole on the third sheet and the $K_0^*(1950)$ pole on the second
sheet, since they do not contribute to the decreasing of total
$\chi^2$.

In recent few years considerable progresses have been made in
studying the low lying scalar resonances in hadron spectrum. For the
$\sigma$ meson, it is demonstrated that the $\sigma$  is crucial for
chiral perturbation theory to accommodate for the CERN--Munich
I,J=0,0 $\pi\pi$ scattering phase sheet data.~\cite{XZ00} A rather
precise determination to the $\sigma$ pole location can be
obtained.~\cite{ACL,ZHOU05} The situation for the $\kappa$ resonance
is less clear, due to the flawed data as well as that crossing
symmetry is not yet used in the determination of the $\kappa$ pole.
Though the current study confirms the existence of the $\kappa$
resonance, more efforts have to be made in order to get a precise
understanding to the $\kappa$ pole location.

 {\it Acknowledgement}:
 We thank one referee's stimulating
 remarks which are helpful in formulating the discussions in its present form.
  This work is support in part by
National
 Nature Science Foundations of China under contract number
 10575002,
 10421503
 and 10491306.


\vspace{.5cm}

\noindent {\bf\large Appendix} \vspace{.5cm}

Here we present the proof of the statement that if the full
scattering amplitude is polynomial bounded then $f(0)=0$. In order
to achieve this, according to the previous text in this paper, one
only needs to prove $T^I_J(s\to 0)$ is no more singular than
$O(s^{-n})$ where $n$ is an arbitrary but finite constant.

The proof can be made following the method of Ref.~\cite{Jakob}, but
in the present case the situation is simpler. From partial wave
projection formula,
 \bqa\label{appendixpartial}
  T^I_J(s)
&=& {1\over 32\pi}{1\over 2{q_s}^2} \int_{-4{q_s}^2}^0
dt P_J(1+{{t}\over {2{q_s}^2}})T^{I}(s,t,u)\ ,
\eqa
  and Eq.~(\ref{auxillary}) we see that,
  \be
  q_s^2\to \frac{(m_K^2-m_\pi^2)^2}{4}s^{-1}\to +\infty \,\,\,\,\, (\mbox{when}\,\,\, s\to 0_+)\ .
   \ee
Hence the asymptotic limit of $T^I_J(s)$ when $s\to 0_+$ is governed
by the asymptotic behavior of $T^I(s,t)$ when $s\to 0_+$ and
$t\to-\infty$, through Eq.~(\ref{appendixpartial}). On the other
side, the asymptotic behavior of $T(s\to 0_-)$ is governed by the
asymptotic behavior of $T^I(s,t)$ when $s\to 0_-$ and $t\to+\infty$.
The physical region of $\pi K\to \pi K$ is given by $t=0$ and the
hyperbola $su=(m_K^2-m_\pi^2)$. The boundaries of the double
spectral regions can be found in, for example, Ref.~\cite{LANG}.
 It is noticed that $T^I(s,t)$ is a regular function of $s$ at $s=0$ for
 any fixed $t$, since on the Mandelstam plot the line $s=0$ does not touch any of the double spectral region.
 Hence we only need to discuss the $t\to \pm\infty$ limit of
  $T^I(s,t)$ for fixed small positive $s$. In such a situation,
  one can prove the following relation,
  \be
\lim_{t\to +\infty}T^I(s,t)=\lim_{t\to -\infty}T^I(s,t)\ \ee
 using a mathematical theorem  which states that: \emph{a function
 which is analytic in the upper half of the complex $z$ plane and
 does not increase exponentially for $|z|\to \infty$ along any ray
 in the upper half of $z$ plane, cannot tend to different limits
 along the positive and negative real axis} (of course only along the upper edge).
The conditions for this theorem to hold are no more than the
Mandelstam analyticity assumption.

Therefore we consider the $s\to 0_+$ limit of
Eq.~(\ref{appendixpartial}), taking $J=0$ only for simplicity.
Assuming $|T(s,t)|<|t|^n$ for any fixed $s$ and large $|t|$, we have
 \bqa\label{pwaTappen2}
\lim_{s\to +0}|T^I_0(s)|& < &\lim_{s\to 0_+} {1\over 32\pi}{1\over
2{q_s}^2} \int_{-4{q_s}^2}^0
dt |T^I(s,t)| \nonumber \\%
& < &\lim_{s\to 0_+} {1\over 32\pi}{1\over 2{q_s}^2}
\int_{-4{q_s}^2}^0
dt |t|^n \nonumber \\%
& \sim & O(s^{-n})\ .
 \eqa
 Hence we complete the proof.

 If higher partial waves are considered, each term of the Legendre
polynomial in Eq.~(\ref{appendixpartial}) contributes at the same
order when $s\to 0$ and the sum results only in an additional
coefficient in front of $O(s^{-n})$. See Ref.~\cite{Jakob} for
details.

\begin{thebibliography}{99}
\bibitem{Bugg}Similar opinion is also emphasized by David Bugg: private
communications.
\end{thebibliography}
\end{document}